# High-temperature Bose-Einstein condensation of polaritons: realization under the intracavity laser pumping of matter condition

V.A. Averchenko[1], A.P. Alodjants[2*], S.M. Arakelian[2], S.N. Bagayev[3], E.A. Vinogradov[4], V.S. Egorov[1], A.I. Stolyarov[1], I.A. Chekhonin[1]

[1] *St. Petersburg State University, Ul'yanovskaya ul. 1, 198504 St. Petersburg, Staryi Peterhof, Russia*
[2] *Vladimir State University, ul. Gor'kogo 87, 600000 Vladimir, Russia*
[3] *Insitute of Laser Physics, Russian Academy of Sciences, prosp. akad. Lavrent'eva 13/3, 630090 Novosibirsk, Russia*
[4] *Institute of Spectroscopy, Russian Academy of Sciences, 142190 Troitsk, Moscow region, Russia*

*Abstract.* **A quantum model of Bose-Einstein condensation based on processes involving polaritons excited in an intracavity absorbing cell with resonance atoms, which is manifested in the spectral characteristics of the system, is considered. It is shown that the spectral 'condensation' appears which is directly related to the degeneracy of a weakly interacting gas of polaritons resulting in quasi-condensation at room temperature. The possibility of obtaining polariton condensation as a new phase state by using the confinement of polaritons in an atomic optical harmonic trap is discussed.**

## 1. Introduction

Experiments on the Bose-Einstein condensation (BEC) of macroscopic numbers of atoms $(N \geq 10^6)$ is one of the most spectacular recent advances, which have made a great influence on the development of various directions in modern quantum and laser physics and newest technologies (see, for example, [1]). In the case of BEC, when under conditions of the temperature phase transition a macroscopic number of atoms are in the ground (lower) quantum level, a new coherent state of matter is formed. This is manifested in the fact that, for example, at the limiting temperature $T=0$ an ensemble of condensate atoms, as each individual atom, is described by the common wave function corresponding to a coherent state. In this aspect, the BEC phenomenon is similar to lasing, for example, when strict phase locking of laser modes occurs in laser cavities [2, 3]. In addition, in the case of BEC, we can say about the realisation of a Bose laser (boser) emitting coherent ensembles of atoms [3, 4]. A remarkable feature of such macroscopic quantum states of matter is the possibility to use them for the development of new physical principles of quantum information processing and communication [5, 6].

However, despite spectacular achievements in this direction, there exist a number of practical difficulties imposing principal restrictions on the possibility of real applications of the atomic BEC for these purposes. Thus, one of the basic problems is the necessity of maintaining extremely low temperatures (tens of nK) to realise such devices. In this connection the problem of obtaining macroscopic coherent (quantum) states of matter at high (room) temperatures becomes very important.

One of the most attractive approaches to the solution of this problem is the preparation of a quasi-condensate of the two-dimensional Bose gas of weakly interacting polaritons (in atomic physics [7]) and excitons (in solid state physics [8-10] [*])). Such collective states of the medium (quasi-particles) represent a superposition of photons and spin waves in the atomic medium and can be obtained, for example, within the framework of the Dicke model used to describe superradiance [12]. Although these states cannot be treated as a condensate in a strict thermodynamic sense due to the nonequilibrium state of the system as a whole, under certain conditions imposed on the type of atomic optical interactions in the system, polaritons do form a condensate, their distribution being described by the Bose-Einstein distribution function for an ideal gas of bosons [13].

_______________________________________________

[*] Email: alodjants@vpti.vladimir.ru

In this paper, we considered the interaction of a system of two-level atoms with an electromagnetic field in the cavity in the case of the so-called strong coupling, when the inequality

$$\omega_c = \left(\frac{2\pi d^2 \omega_0 n}{\hbar}\right)^{1/2} \gg \frac{1}{2\tau_{coh}}, \qquad (1)$$

is fulfilled, where $\omega_c$ is the cooperative frequency determining the collective interaction of atoms with the field; $\omega_0$ is the atomic transition frequency; $d$ is the transition dipole moment; $\tau_{coh}$ is the characteristic coherence time of the atomic medium; $n$ is the atomic gas density; and $\hbar$ is Planck's constant. In this case, the field itself is weak (in the number of photons).

The so-called condensation of the spectrum occurs when inequality (1) is fulfilled [14, 15]. This effect consists in the fact that under some threshold conditions imposed on the concentration of absorbing atoms and pump intensity, radiation of a broadband laser with a narrowband absorbing intracavity cell is concentrated ('condensed') near the strongest absorption lines of matter.

This phenomenon was observed experimentally and interpreted within the framework of a clear classical model of parametric excitation of two coupled oscillators (electromagnetic field and atoms of matter) upon coherent energy transfer between them. However, this model is one-dimensional and is not quantum one, which obviously restricts the field of its applications.

In this paper, we propose a detailed quantum model of spectral condensation realised for polaritons excited in an intracavity absorbing cell [16]. We show that spectral condensation can be directly related to the condensation (quasi-condensation) of polaritons in the cavity if a strong coupling between the electromagnetic field and medium is provided. The latter statement is in itself of interest, and in this paper we substantiate for the first time the possibility of obtaining the true BEC in the polariton system at high (room) temperatures in the case of spectral condensation. In this respect, of interest are the experimental data [8, 9] obtained in semiconductor microcavities, which confirm the above assumption.

## 2. Basic relations

Consider the interaction of two-level atoms (with levels *a* and *b)* with a quantum electromagnetic field, which is described by the photon annihilation (creation) operators $f_k (f_k^+)$ for the *k*-th mode. Within the framework of dipole approximation, such a system can be described by the Hamiltonian [13]

$$H(k) = \sum_k E_{ph}(k) f_k^+ f_k + \sum_k \tfrac{1}{2} E_{at}(b_k^+ b_k - a_k^+ a_k) + \sum_k g(f_k^+ a_k^+ b_k - b_k^+ a_k f_k), \qquad (2)$$

where $a$ ($a^+$) and $b$ ($b^+$) are the Bose operators of annihilation (creation) of atoms at the lower and upper levels, respectively; $E_{ph}(k) = \hbar c |\vec{k}|$ is the dispersion relation for the photons in the resonator; $E_{at} \equiv E_0$ is the transition energy in a two-level atom (here we neglect the motion of atoms in the resonator); $g$ is the coefficient determining the strength of coupling between the field and an atom; and $c$ is speed of light in vacuum.

In the limiting case of small perturbations of the atomic system, atoms occupy mainly the lower energy level *a,* and the inequality

---

[*]) Here, we are dealing with the so-called Kosterlitz-Thouless phase transition to the superfluid state of two-dimensional Bose systems in which the true Bose-Einstein condensation (in the absence of confinement of gas particles in a trap) is impossible [11].

$$\langle b_k^+ b_k \rangle \ll \langle a_k^+ a_k \rangle \qquad (3)$$

is fulfilled. In this approximation, Hamiltonian (2) can be diagonalised by using the unitary transformation

$$\Phi_{1,k} = \mu_k f_k - \nu_k a_k^+ b_k, \qquad \Phi_{2,k} = \nu_k f_k + \mu_k a_k^+ b_k, \qquad (4)$$

where the introduced annihilation operators $\Phi_{j,k}$ (j=1,2) characterise quasi-particles (polaritons) in the atomic medium, corresponding to two types of elementary perturbations, which in approximation (3) satisfy the boson commutation relations

$$\left[\Phi_{i,k}; \Phi_{j,k}^+\right] = \delta_{ij}, \quad i,j = 1,2. \qquad (5)$$

The transformation parameters $\mu_k$ and $\nu_k$ in expression (4) are real Hopfield coefficients satisfying the condition $\mu_k^2 + \nu_k^2 = 1$, which determine the contributions of the photon and atomic (excited) components to a polariton, respectively:

$$\mu_k^2 = \frac{4g^2}{2(\delta_k^2 + 4g^2)^{1/2}\left[\delta_k + (\delta_k^2 + 4g^2)^{1/2}\right]}, \qquad \nu_k^2 = \frac{\delta_k + (\delta_k^2 + 4g^2)^{1/2}}{2(\delta_k^2 + 4g^2)^{1/2}}, \qquad (6a,b)$$

where $\delta_k = E_{at} - E_{ph}(k)$ is the phase mismatch determining the contributions of the photon and atomic components to expression (4) for polaritons. In particular, in the limiting case, when $\delta_k \ll -2|g|$, we have $\mu_k^2 \to 1$ ($\nu_k^2 \to 0$), which corresponds to the negligible contribution of the photon part to the polariton $\Phi_{2,k}$. In opposite limit, when $2|g| \ll \delta_k$, we have $\mu_k^2 \to 0$ ($\nu_k^2 \to 1$), which means that the photon contribution to the coherence of polaritons of this type increases. Expression (6) shows that the polariton is a half-matter and half-photon ($\mu^2 = \nu^2 = 1/2$) quasi-particle under the resonance condition $\delta_k = 0$.

Taking expressions (4) and (6) into account, Hamiltonian (2) takes the form

$$H(k) = \sum_k E_1(k)\Phi_{1,k}^+\Phi_{1,k} + \sum_k E_2(k)\Phi_{2,k}^+\Phi_{2,k}, \qquad (7)$$

where $E_{1,2}(k)$ determine the dispersion dependence of polaritons:

$$E_{1,2}(k) = \frac{1}{2}\left\{E_{at} + E_{ph}(k) \pm \left\{\left[E_{at} - E_{ph}(k)\right]^2 + 4g^2\right\}^{1/2}\right\}. \qquad (8)$$

Figure 1a presents dispersion dependences $E_{1,2}(k)$ (8) of polaritons for the interaction of atoms with the quantum field in free space. One can see that the two allowed energy states, polaritons of the upper [$E_1(k)$] and lower [$E_2(k)$] branches, correspond to each value of the wave vector $k$.

When the medium is placed into the resonator, the wave-vector component $k_\perp$ orthogonal to the mirror surface is quantised. At the same time, a continuum of modes exists in the direction parallel to the mirror surface due to the absence of boundary conditions. This means that in the single-mode (single-frequency for each value of $k_\perp$) regime, the dispersion of polaritons is determined only by the wave-vector component $k_\parallel$ parallel to the mirror surface. Then, under the condition $k_\parallel \ll k_\perp$ which corresponds physically to the paraxial approximation in optics (see, for example, [17]), the dispersion relation for photons in the resonator has the form

$$E_{ph}(k) = \hbar c(k_\perp^2 + k_\parallel^2)^{1/2} = \hbar c\left[k_\perp + \frac{k_\parallel^2}{2k_\perp} + O\left(\frac{k_\parallel^3}{k_\perp^2}\right)\right]. \qquad (9)$$

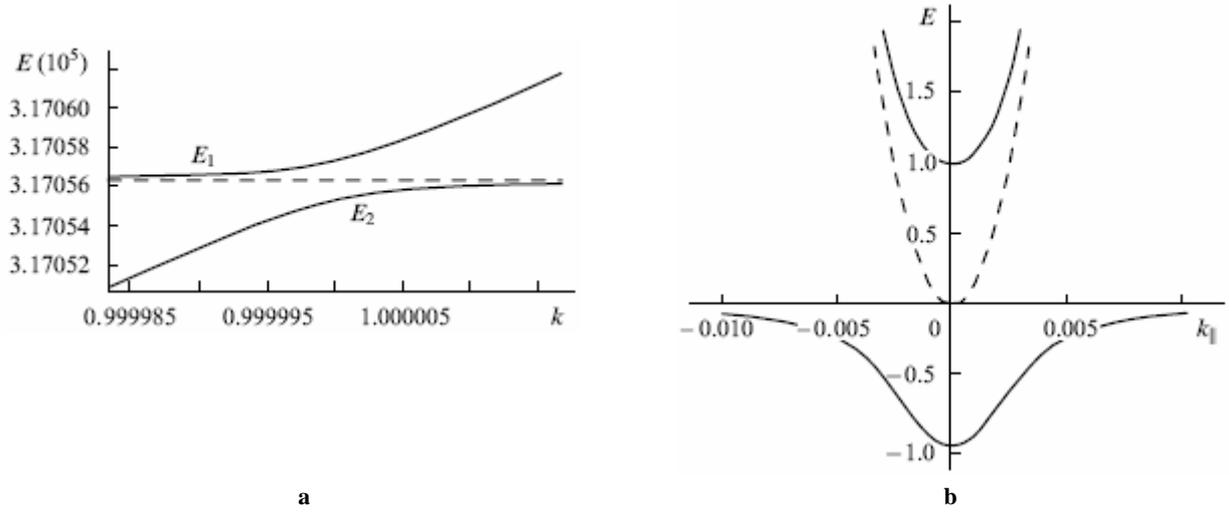

**Figure 1.** Dispersion dependences $E_1(k)$ (upper branch) and $E_2(k)$ (lower branch) of polaritons on the wave vector $k$ in free space (a) and resonator (b). The wave vector is plotted on the abscissa in the units of the resonance wave vector $k_\perp$ on the ordinate the energy is plotted in the units of the coupling coefficient $g$.

Here $k_\perp = \pi m / L_{cav}$ is the quantised component of the wave vector parallel to the resonator axis, which corresponds to the periodic boundary conditions in the standard field quantisation procedure; $L_{cav}$ is the effective resonator length; and the number $m$ corresponds to the selected mode (frequency). In the case of strong coupling (1), the dispersion curves of a polariton are pushed apart, resulting in the appearance of the upper and lower polariton branches in the resonator (Fig. 1b). The principal feature of these curves is the presence of the 'potential' well (for $k = 0$). The width of the lower polariton well can be found from the condition $\partial^2 E_2 / \partial k_\parallel^2 = 0$. This condition determines the angular parameters of a polariton beam in the resonator. It is important to note that these effects, which are related to the transverse component of the wave vector of a polariton ($k_\parallel$ in our case), will not be suppressed due to light diffraction if the angular dimensions of the polariton beam exceed the diffraction-limited divergence $\varphi$ of the light beam, which can be estimated from the expression $\varphi \approx d / L_{cav}$ [$d$ and $L_{cav}$ are the beam diameter and resonator (or absorbing cell) length, respectively].

## 3. Spectral 'condensation' and condensation of polaritons

Within the framework of our approach, the narrowing ('condensation') of the polariton spectrum, which was observed in experiments [14, 15], can be simply explained by BEC. In this connection, taking into account paraxial approximation (9), we represent Hamiltonian (7) in the form

$$H = H_{long} + H_{tr}, \tag{10a}$$

where

$$H_{long} = \sum_{k_\perp} E_1'(k_\perp) \Phi_{1,k_\perp}^+ \Phi_{1,k_\perp} + \sum_{k_\perp} E_2'(k_\perp) \Phi_{2,k_\perp}^+ \Phi_{2,k_\perp}, \tag{10b}$$

$$H_{tr} = \sum_{k_\parallel} E_{1,tr}'(k_\parallel) \Phi_{1,k_\parallel}^+ \Phi_{1,k_\parallel} + \sum_{k_\parallel} E_{2,tr}'(k_\parallel) \Phi_{2,k_\parallel}^+ \Phi_{2,k_\parallel} \tag{10c}$$

The expression for $H_{long}$ describes polaritons formed along the resonator axis, $E_{1,2}'(k_\perp) \equiv E_{1,2}'(k)\big|_{k_\parallel = 0}$ determines their dispersion dependence [see (8)] for $k_\parallel = 0$. The expression for $H_{tr}$ characterises polaritons produced in the two-dimensional plane perpendicular to the resonator axis. The dispersion of these polaritons is described by the expression $E_{1,2,tr}' = \tfrac{1}{2} \hbar^2 k_\parallel^2 / m_{pol}^{(1,2)}$. Here,

$$m_{pol}^{(1,2)} = \frac{2m_{ph}}{1 \mp \Delta/(\Delta^2 + 4g^2)^{1/2}} \qquad (11)$$

is the mass of polaritons of the upper and lower branches; $m_{ph} = \hbar k_\perp / c \approx E_0/c^2$ is the effective photon mass in the medium and $\Delta = E_0 - \hbar c k_\perp$ is the detuning of the resonator mode (frequency) from the atomic transition frequency.

Thus, the BEC of polaritons in the resonator is related to the second term in the expression for the Hamiltonian $H$ in (10a). This term leads in fact to the renormalisation of the photon mass in the medium [see (11)]. Quasi-particles (polaritons) appearing in this case can be treated as an ideal two-dimensional gas [see also (10b)]. Indeed, the possibility of BEC assumes the presence of a stable state with the minimal energy - a 'potential' well (at the point $k_\parallel = 0$), which, as shown in section 2, takes place for polaritons in the resonator (the lower branch in Fig. 1b).

The resonator parameters can be selected simultaneously so that the equality $|k_\perp| = E_0/\hbar c$ would be fulfilled, i.e., $\Delta = 0$. Thus, polaritons can condense in the resonator to the state in which the resonance interaction of the field and matter is not destroyed. Therefore, BEC increases the effective $Q$ factor of the resonator, thereby reducing the threshold pump power, which should be exceeded to produce spectral condensation in the model of parametric interaction of coupled oscillators.

By using expressions (8), (9), and (11) for $E_2(k)$ we obtain that the width (and depth) of the polariton well expressed in energy units is of the order of the coupling coefficient

$$\frac{\hbar^2 \Delta k_\parallel^2}{2m_{eff}} \approx g. \qquad (12)$$

In this case, it is possible to introduce formally the effective temperature $T_{eff}$ of the two-dimensional Bose gas of polaritons, which is also of the order of the coupling coefficient within the polariton well [13], i.e., $K_B T_{eff} \approx g$, where $K_B$ is the Boltzmann constant.

The approach discussed above determines the condensation (more exactly, quasi-condensation) of the two-dimensional gas by assuming that polaritons with large $k_\parallel$ efficiently relax to the bottom of the dispersion-curve well. In our case, unlike the case of semiconductor microcavities considered in [8, 9], the two-dimensional property of the polariton gas can be provided by the fact that an optically dense medium is excited, as a rule, by the wave packet of synchronised electromagnetic modes, which corresponds to the quasi-monochromatic interaction of the field with medium.

The efficient relaxation of polaritons to the bottom of the 'dispersion' well can be related to the intense polariton-polariton interaction discussed in a number of papers (mainly concerning the problems with semiconductor micro-cavities [10, 18]).

Consider now in more detail the quasi-condensation of a two-dimensional Bose gas of polaritons described by the last term in (10c). The chemical potential of such a gas is described by the expression [19]

$$\mu = K_B T \ln\left[1 - \exp(-n_2 \lambda_T^2)\right] \equiv K_B T \ln\left[1 - \exp\left(-\frac{T}{T_d}\right)\right], \qquad (13)$$

where $T_d = 2\pi\hbar^2 n_2/(m_{eff} K_B)$ is the gas degeneracy temperature; $n_2$ is the two-dimensional density of polaritons in the plane perpendicular to the resonator axis; and $\lambda_T = \hbar/(2m_{eff} K_B T)^{1/2}$ is the thermal wavelength (de Broglie wavelength). The temperature $T_d$ in (13) is determined by the condition when the thermal wavelength $\lambda_T$ is of the order of the average distance $V^{1/3}$ between particles ($V$ is the system volume). Due to the interaction between polaritons (nonideal gas), the additional parameter $a_{scat}$ appears, which is the scattering length depending on the interaction potential. This parameter affects the energy spectrum, which becomes a phonon spectrum [9].

It follows from (13) that, strictly speaking, the condensation of the two-dimensional polariton gas ($\mu = 0$) occurs at $T \to 0$. At the same time, it is known (see, for example, [9]) that already at the temperature

$$T_{KT} = \frac{\pi \hbar^2 n_s}{2 m_{eff} K_B} \simeq \frac{T_d}{4} \tag{14}$$

the Kosterlitz-Thouless phase transition to the superfluid state occurs in the two-dimensional weakly interacting Bose gas, when isolated condensate droplets with uncorrelated phases are formed on the two-dimensional surface [$n_s$ in (14) is the superfluid liquid density on the two-dimensional surface].

For polaritons with the effective mass $m_{eff} = 5 \times 10^{-33} g$ and density in a three-dimensional resonator $n_3 = 3.5 \times 10^{11} cm^{-3}$, the gas degeneracy can appear already at room temperature ($T_d$ = 300 K). Indeed, in this case the minimal two-dimensional density of the polariton gas estimated from (13) for $T = T_d$ gives the value $n_2 \simeq \lambda_T n_3 \approx 0.3 \times 10^8 cm^{-2}$ [19] for the de Broglie wavelength $\lambda_T \approx 1.84 \times 10^{-4} cm$. It is for this value of the atomic concentration $n_3$ that spectral condensation was observed near the yellow doublet of sodium in experiments [14] (Fig. 2). For the upper spectrum (Fig. 2a), $n_3 < 10^{10} cm^{-3}$, and for the lower spectrum (Fig. 2c), $n_3 = 3.5 \times 10^{11} cm^{-3}$. The similar results obtained in [14] for the neon spectrum also demonstrated the spectral condensation for polaritons.

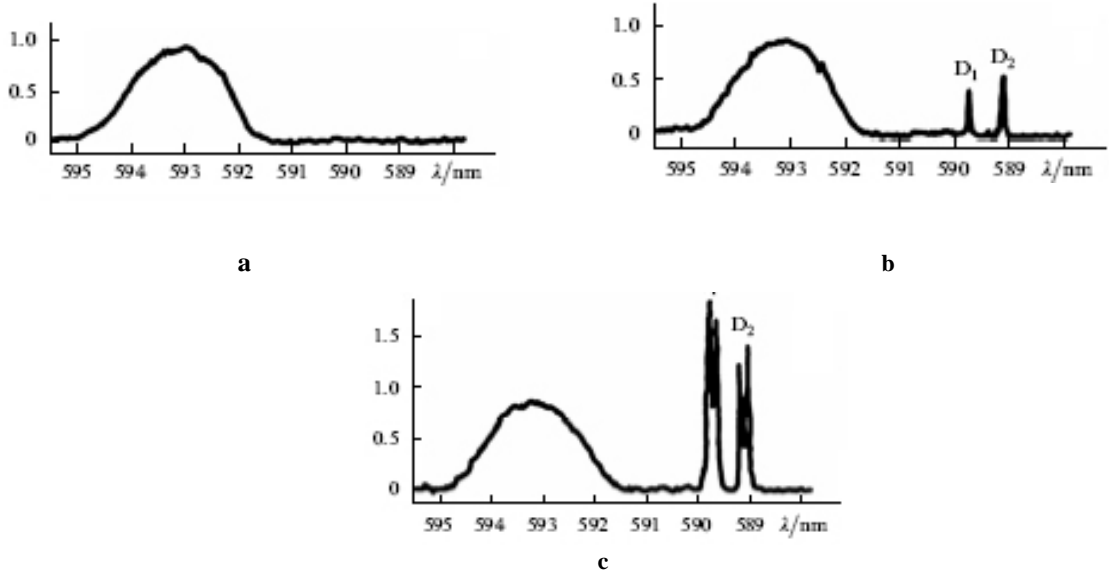

**Figure 2.** Spectral condensation near the yellow doublet of sodium (see text) at the atomic concentration $n_3 \leq 10^{10}$ (a), $10^{11}$ (b), and $3.5 \times 10^{11} cm^{-3}$ (c).

Let us find now the conditions under which the true (in thermodynamic sense) condensation of polaritons excited in the resonator can be obtained. It is known [19, 20] that, to obtain such condensation in a two-dimensional weakly interacting (ideal) gas, gas particles should be confined in a trap. For example, for a trap with the trapping potential described by the expression (harmonic potential)

$$U(r) = U_0 \frac{r^2}{r_0^2} = \frac{m_{eff} \Omega_{eff}^2}{2} r^2 \tag{15}$$

($\Omega_{eff}$ is the trapping (oscillation) frequency of particles, $r_0$ is the transverse size of the trapping region, and $r$ is the transverse coordinate), the critical BEC temperature for a two-dimensional gas is (cf. [20])

$$T_c = \frac{\hbar \Omega_{eff}}{K_B} \sqrt{\frac{N}{1.645}} = \frac{2\pi \hbar^2 n_2}{1.645 m_{eff} K_B}, \qquad (16)$$

where $N$ is the total number of particles. We also took into account in the right-hand side of (16) that the number $N_2$ of particles trapped by the potential $U(r)$ on the surface is described by the expression [20] $N_2 = 2\pi n_2 K_B T / m_{eff} \Omega_{eff}^2$. In the absence of a trap ($\Omega_{eff} = 0$), as should be, BEC does not occur: $T_c = 0$ in (16). The scheme of an atomic optical trap for confinement of polaritons is discussed in Appendix.

Let us make the estimates for a polariton gas with the number density of particle $n_3 = 3.5 \times 10^{11} cm^{-3}$, which corresponds to the two-dimensional concentration $n_2 = 0.5 \times 10^8 cm^{-2}$. In this case, condensation in a trap can also occur at room temperature ($T_c$ = 300 K). These values of the gas density, as follows from Fig. 2c, correspond to the conditions of experiments [14, 15], in which spectral condensation was observed. It is interesting to note that in this case the estimation $\lambda_T \geq r_{int}$ for the critical de Broglie wavelength is valid ($r_{int} \sim 1/\sqrt{n_2} \approx 1.41 \times 10^{-4} cm$ is the characteristic scale of the distance between particles on the surface). This means in fact that the wave functions of polaritons overlap with each other. As a result, the two-dimensional polariton gas at temperatures $T \leq T_c$ confined in a harmonic trap (15) is in the ground state. The number $N_0$ of particles in this state depends on temperature as $N_0 \approx N \left[ 1 - (T/T_c)^2 \right]$. Thus, macroscopic BEC occurs already at room temperature. Such a state is coherent and polaritons have the same phases. Even in the presence of some phase mismatches, the individual wave functions of polaritons can be probably correlated with the help of a rather weak external (stabilizing) electromagnetic field.

In this paper, we did not consider spatial effects and the problems of stability of polariton BEC in a trap. The experimental study of the transverse coherence effects, the superfluidity of a polariton gas, and the influence of the trap potential on the dynamics of the condensate 'cloud' is very important and requires a separate analysis. We point out here only the following substantial circumstance.

At present, the experimental investigation of quantum (coherent) properties of polaritons in resonators (measurements of the first- and second-order coherence degree) is one of the main tools for diagnostics of polariton condensation (see [8, 9]). Because a polariton is a linear superposition of a photon and atomic excitation [see (4)], its coherent properties are caused by the coherence of the light field itself and of an ensemble of atoms with which the field interacts, as well as by their possible quantum interference caused by the condensation process. Within the framework of these experiments, when the condition of the exact resonance $\Delta = 0$ is fulfilled, we have $\delta_k \approx 0$ and obtain $\mu_k^2 = v_k^2 = 1/2$ from expressions (6) and (11), which means that optical and atomic parts make identical contributions to a polariton. In this case, the coherent properties of the polariton state can be simply caused by a high coherence of the optical field at the input to the atomic medium irrespective of BEC. However, the problem of measuring the coherence of atomic exitations caused by the interaction and of the intrinsic coherence of the polariton condensate (if it is produced in the system) remains open. In our opinion, this problem can be solved, in particular, by producing polariton BEC based on three-level atoms under conditions of electromagnetic induced transparency (EIT) (see below).

Here we consider another possibility based on a small variation of the detuning $\Delta$ [and, therefore, $\delta_k$, see (6)] in experiments as the parameter governing the contributions of photon and atomic parts to the resulting coherence of resonator polaritons. In this case, the effective mass of polaritons [see (11)] and, hence, the critical temperatures of degeneracy, condensation, and quasi-condensation in (13), (14), and (16) change. This specific property of a polariton gas means in fact that the formation of a Bose-Einstein condensate can be controlled in experiments.

Note, however, that we do not consider in this paper the questions concerning the BEC of a photonic gas in the resonator or, more exactly, the condensation of polaritons of the upper branch of the dispersion curve (see Fig. 1b) characterised by the first term in expression (10c). This problem is undoubtedly very important for the scope of questions considered in our paper although it was discussed only in connection with the quantum properties of light in media with cubic nonlinearity (see [3]).

In addition, the formation of a photon condensate (or a condensate of polaritons of the upper branch), which is directly connected with lasing in the resonator (cf. [2]), upon varying the parameter $\Delta$ also has an interesting feature. Indeed, for $\Delta = 0$, it follows from (11) that polaritons of both dispersion branches have equal masses, i.e., $m_{pol}^{(1)} = m_{pol}^{(2)} = m_{eff}$, which corresponds to the equal temperatures of their quasi-condensation [see (14)]1. However, in the case of $\Delta \neq 0$, we have from (11) that $m_{pol}^{(1)} \neq m_{pol}^{(2)}$ which means physically that the phase-transition temperatures (14) for polaritons of the upper $(T_{KT}^{(1)})$ and lower $(T_{KT}^{(2)})$ dispersion branches are different. Thus, by introducing asymmetry with the help of a small change in the detuning $\Delta$, it is possible to produce a very narrow temperature (energy) gap within which the coherent properties of polaritons of both branches should substantially change. These properties can be observed, for example, by measuring the function of their cross correlation or by using probe radiation under resonance conditions.

Therefore, the study of this effect will give the answer to the principal question about the properties of the coherence of light, atomic system, and polaritons themselves in the case of BEC.

## 4. Conclusions

We have developed in the paper the quantum approach for solving the problems of formation of quasi-condensation and realisation of the true (in the thermodynamic sense) Bose - Einstein condensation of a two-dimensional gas of polaritons at room temperature. This approach has allowed us to explain some features of spectral condensation of broadband lasing near strong absorption lines in the laser resonator, which were observed in experiments (in particular, the so-called spectral condensation upon non-resonance pumping). In this aspect, BEC reduces the threshold pump power of parametric excitation of cooperative effects. Consider briefly some phenomena that are directly related to the problem studied in the paper.

First, this is the condensation of polaritons, which is of interest in the presence of the EIT effect when a light pulse propagates in a resonance atomic medium without changing its shape in the absence of absorption (see, for example, [5, 21, 22]. A remarkable feature of this effect is the appearance of atomic coherence both for hot [21] and ultracold atoms [5, 22]. The EIT effect can be also explained in terms of bright and dark polaritons, which in the adiabatic approximation corresponding to condition (3) in our case, represent the coherent superposition of atoms in the two states of the hyperfine Zeeman structure and the external probe field maintained with the help of the external probe field at the optical frequency through the third (auxiliary) level (the so-called $\Lambda$ - scheme [5, 6, 21]).

Therefore, upon placing an atomic medium into the resonator to produce the BEC of polaritons, the EIT effect would become a tool for obtaining such a quantum state. In this case, the condensation process could be controlled more precisely by coupling directly two atomic levels with an external weak field, which would provide the ejection of 'hot' polaritons from a trap, as, for example, occurs for condensation of alkali atoms in a magneto-optical trap [1]. On the other hand, upon spectral condensation in the case of BEC, a 'bleaching' of the atomic medium in the resonator caused by a change in its refractive properties can be expected. In this case, the group velocity of a light pulse directed into an atomic medium after switching on probe radiation with the delay time $\tau_{del} < \tau_{coh}$ can decrease, in particular, due to polariton condensation. Indeed, it follows from expressions (10c) and (11) that the group velocities of such quasi-particles in the plane perpendicular to the resonator axis are determined by the expression

$$\upsilon_{gr}^{(1,2)} = \frac{\partial E_{1,2tr}}{\partial (\hbar k_\parallel)} = \frac{\hbar k_\parallel}{m_{pol}^{(1,2)}}$$

In the case of the exact atomic optical resonance (for $\Delta = 0$), we have from this that $\upsilon_{gr}^{(1)} \simeq \upsilon_{gr}^{(2)} = k_\parallel c / 2k_\perp$ Therefore, in the paraxial approximation, when $k_\parallel \ll k_\perp$ the group velocity of condensed polaritons is estimated as $\upsilon_{gr}^{(1,2)} \ll c$, which means in fact that the 'slow' light regime is observed for polaritons in the resonator.

Second, the high-temperature BEC of polaritons is of interest for quantum information, for example, for the development of new physical principles of quantum memory and data storage. Indeed, as we have shown in [6], such macroscopic polariton states can be used in problems of cloning and quantum information storage.

*Acknowledgements.* This work was partially supported by the Russian Foundation for Basic Research (Grants Nos 04-02-17359 and 05-02-16576) and the Ministry of Education and Science of the Russian Federation. A.P. Alodjants thanks the non-profit Dynasty Foundation for support.

# Appendix

Let us discuss the problem of confinement of the BEC of intracavity polaritons in a trap. Consider a special atomic optical trap whose operation is based on the fact that polaritons represent a coherent superposition of a photon and atomic perturbation. Photons can be confined in the region of atomic-optical interaction in such a trap, where polaritons are produced, by focusing a light beam with a special gradient (cylindrical) lens (or inhomogeneous waveguide) with the refractive index varying along the transverse coordinate as

$$n^2(r) = n_0^2(1 - n'r^2), \qquad (A.1)$$

where $n'$ is the required gradient addition to the refractive index of the lens. The potential for trapping (focusing) photons of the light beam produced by such an optical system can be written in the form [17]

$$U_{opt}(r) = \frac{n^2(r) - n_0^2}{2n_0^2} = \frac{n'r^2}{2},$$

which exactly corresponds to the harmonic-trap potential (15) with the inhomogeneity parameter $n' = m_{eff}\Omega_{eff}^2$.

In addition, to trap atoms in the plane perpendicular to the resonator axis, we can use a two-dimensional magnetic trap with the oscillation frequency $\Omega_{at}$, which is widely applied in experiments with 'usual' atomic condensates [1].

Thus, to confine polaritons in a trap, it is necessary to confine atoms by a standard method and focus simultaneously the light beam into the region of atomic-optical interaction by selecting the appropriate parameters $\Omega_{at}$ and $n'$. This determines the value of $\Omega_{eff}$ required in the experiment.

# References

1. Ketterle V. *Usp. Fiz. Nauk,* **173,** 1339 (2003).
2. Oraevsky A.N. *Kvantovaya Elektron.,* **24,** 1127 (1997) [*Quantum Electron.,* **27,** 1094 (1997)].
3. Chiao R., Boyce J. *Phys. Rev. A,* **60,** 4114 (1999).
4. Imamoglu A., Ram R.J., Pau S., Yamamoto Y. *Phys. Rev. A,* **53,** 4250 (1996).
5. Liu C, Dutton Z., Behroozi C.H., Hau L.N. *Nature,* **409,** 490 (2001).
6. Alodjants A.P, Arakelian S.M. *Int. J. Mod. Phys. B,* **20,** 1593 (2006).
7. Averchenko V.A., Bagayev S.N., et al. *Abstract in Technical Digest o/ICONO'05 Conf.* (Sankt-Petersburg, Russia, 2005).
8. Deng H., Weihs G., Santori C, Bloch J., Yamamoto Y. *Science,* **298,** 199 (2002).
9. Kavokin A., Malpuech G., Laussy F.P. *Phys. Lett. A,* **306,** 187 (2003); Richard M., Kasprzak J., Andre R., et al. *Phys. Rev. B,* **72,** 201301(R) (2005).
10. Gippius N.A., Tikhodeev S.G., Keldysh L.V., Kulakovskii V.D., *Usp. Fiz. Nauk,* **175,** 327 (2005); Kulakovskii V.D., Krzhizhanovskii D.N., et al. *Usp. Fiz. Nauk,* **175,** 334 (2005).
11. Kosterlitz J.M., Thouless D.J. *J. Phys. B: Sol. State Phys.,* **6,** 1181 (1973).
12. Dicke R.H. *Phys.Rev.,* **93,** 99 (1954).
13. Eastham P.R., Littlewood P.B. *Phys. Rev. B,* **64,** 235101 (2001).
14. Vasil'ev V.V., Egorov V.S., Fedorov A.N., Chekhonon LA. *Opt. Spektr.,* **76,** 146 (1994).
15. Bagayev S.N., Egorov V.S., Moroshkin P.V., Fedorov A.N., Chekhonon LA. *Opt. Spektr.,* **86,** 912 (1999).
16. Kocharovskii V.V., Kocharovskii Vl.V. *Kvantovaya Elektron.,* **14,** 2246 (1987) [*Sov. J. Quantum Electron.,* **17,** 1430 (1987)].
17. Marte M.A., Stenholm S. *Phys. Rev. A,* **56,** 2940 (1997).
18. Savvidis P.G., Baumberg J.J., Stevenson P.M., et al. *Phys. Rev. Lett.,* **84,** 1547 (2000).
19. Petrov D.S., Gangardt G.M., Shlyapnikov G.V. *J. Phys. IV France,* **116,** 3 (2004).
20. Bagnato V., Kleppner D.K. *Phys. Rev. A,* **44,** 7439 (1991).
21. Lukin M.D. *Rev. Mod. Phys.,* **75,** 457 (2003).
22. Prokhorov A.V., Alodjants A.P., Arakelyan S.M. *Pis'ma Zh. Eksp. Tear. Fiz.,* **80,** 870 (2004).